\documentclass[twoside,11pt]{article}
\usepackage{cite,amsfonts,amssymb,exscale,bbm,a4wide,epsfig}
\newcommand{\cl}[1]{\mbox{${\cal #1}$}}

\newcommand{\color}[2][c]{}

\def\Z{{\mathbbm Z}}

\newcommand {\be}{\begin{equation}}
\newcommand {\e}{\end{equation}}
\newcommand {\bea}{\begin{eqnarray}}
\newcommand {\ea}{\end{eqnarray}}
\newcommand {\nit}{\noindent}

\newcommand {\bit}{\bibitem}

\newcommand {\alga}{{\mathfrak a}}
\newcommand {\algb}{{\mathfrak b}}
\newcommand {\algc}{{\mathfrak c}}
\newcommand {\algd}{{\mathfrak d}}
\newcommand {\alge}{{\mathfrak e}}
\newcommand {\f}{{\mathfrak f}}
\newcommand {\g}{{\mathfrak g}}

\newcommand {\fract}[2]{\mbox{${\textstyle{\frac{#1}{#2}}}$}}

\def\ie{{\sl i.e.\/}}
\def\eg{{\sl e.g.\/}}

\def\dopreprint{\hfill{\small\thepreprint}\\}%
\def\preprint#1{\def\thepreprint{#1}}%
\def\thepreprint#1{}%
\def\address#1{\date{{\sl #1}\\\ \\\theversion}\gdef\date##1{}}%
\def\version#1{\gdef\theversion{#1}}%

\version{10 January 2003}%
\preprint{DAMTP-2002-90}

\begin{document}
%
%==============================================================================

\title{\dopreprint Representations of the exceptional and other Lie algebras\\
       with integral eigenvalues of the Casimir operator}
\author{A.~J.~Macfarlane${}^1$\thanks{e-mail: A.J.Macfarlane@damtp.cam.ac.uk}\ \ and
        Hendryk Pfeiffer${}^{2,3,1}$\thanks{e-mail: H.Pfeiffer@damtp.cam.ac.uk}}
\address{\small ${}^1$ Centre for Mathematical Sciences, DAMTP, Wilberforce Road, Cambridge CB3 0WA, UK\\
         ${}^2$ Perimeter Institute for Theoretical Physics, 35 King Street N, Waterloo ON, N2J 2W9, Canada\\
         ${}^3$ Emmanuel College, St.~Andrew's Street, Cambridge CB2 3AP, UK}

\vspace{1cm}
\date{\version}

\maketitle

\begin{abstract}

The uniformity, for the family of exceptional Lie algebras $\g$, of
the decompositions of the powers of their adjoint representations is
well-known now for powers up to the fourth. The paper describes an
extension of this uniformity for the totally antisymmetrised $n$-th
powers up to $n=9$, identifying (see Tables~\ref{tab_order5}
and~\ref{tab_order69}) families of representations with integer
eigenvalues $5,\ldots,9$ for the quadratic Casimir operator, in each
case providing a formula (see eq.~(\ref{ab5}) to~(\ref{ab9})) for the
dimensions of the representations in the family as a function of $D=
\dim \; \g$. This generalises previous results for powers $j$ and
Casimir eigenvalues $j$, $j \leq 4$. Many intriguing, perhaps
puzzling, features of the dimension formulas are discussed and the
possibility that they may be valid for a wider class of not
necessarily simple Lie algebras is considered.

\end{abstract}

% ============================================================================================
%
\section{Introduction}
%
% ============================================================================================

After noting some conventions in Sec.~\ref{sect_notation}, we describe
quite carefully the context of this paper, in
Sec.~\ref{sect_context}. This enables us to indicate briefly in
Sec.~\ref{sect_content} the scope of this paper, and highlight the new
results it obtains.

%--------------------------------------------------------------------------------------------
\subsection{Notation and conventions}
%--------------------------------------------------------------------------------------------
\label{sect_notation}

We are concerned with simple complex Lie algebras $\g$ and with their
irreducible representations. Irreducibility is understood over the
field of complex numbers. We note that we use the informal
abbreviation irrep for irreducible representation.

We focus in particular on the family of 
\be
\label{I1} \cl{F}=\{ \alga_1, \alga_2, \g_2, \algd_4, \f_4, \alge_6, \alge_7, \alge_8 \}
\e 
\nit of simple Lie algebras. They feature in the extension of the last
line of the Freudenthal magic square \cite{freu} that is given in
\cite{ff}. These algebras are well-known to form a family in some profound
sense whose ramifications probably have not yet been fully exhausted.

Our work depends heavily on access to a large body of data for the Lie
algebras $\g$, especially lists for the exceptional Lie algebras of
irreducible representations $R$ classified by highest weights which
give the corresponding dimension and eigenvalue $c^{(2)}(R)$ of the
quadratic Casimir operator ${\cal C}^{(2)}(R)$. We have created a C++
program to provide this and related information given the Cartan
matrix as the only input. We note also that valuable general sources
of data regarding Lie algebras are available, \eg\ \cite{FSS} and
\cite{slan}.

We use a normalisation in which $c^{(2)}(\g)=1$ for the adjoint
representation and therefore
$c^{(2)}(R)=\left<\Lambda_R,\Lambda_R+2\delta\right>$ where $\Lambda_R$
denotes the highest weight of $R$, $\delta$ is the half-sum of
positive roots of $\g$ and $\left<\cdot,\cdot\right>$ denotes the
Cartan--Killing form on the space of weights.

We refer to irreps here often by their dimension because our studies
are concerned with dimension formulas for families of representations
of Lie algebras $\g$. When we need to refer to irreps by their highest
weight or Dynkin coordinate specification, we adopt the conventions
that follow from the Cartan matrices of $\g$ used by
\cite{FSS,slan,cor}. We also often omit commas between the
coordinates, here always integers less than ten.

The diagram automorphisms of the algebras $\g\in\cl{F}$ are $\Z_2$ for
$\alga_2$ and $\alge_6$, $S_3$ for $\algd_4$, and the trivial group for
all the others. As the adjoint irrep $ad$ is always mapped to itself
under diagram automorphisms, the constituents of the complete
decomposition of its tensor powers $ad^{\otimes j}$ are either
self-conjugate or pairs of complex conjugate irreps for $\alga_2$ and
$\alge_6$. For $\algd_4$, the constituents are either irreps that are
stable under triality or triples and sextuples of irreps that are
related by triality.

%--------------------------------------------------------------------------------------------
\subsection{Background material}
%--------------------------------------------------------------------------------------------
\label{sect_context}

The first property of the family $\cl{F}$ to be noted concerns the
structure of $ad \otimes ad$. We write this as
\be \label{I2}
ad \otimes ad = (ad \otimes ad)_A \oplus (ad \otimes ad)_S . \e \nit 
For the antisymmetric piece we have a universal result, \ie\ one
that is valid for each simple compact $\g$
\be \label{I3}
(ad \otimes ad)_A = ad \oplus X_2 , 
\e 
\nit where $X_2$ denotes a representation of $\g$ of dimension 
\be
\label{I4} \dim \; X_2=\fract{1}{2} D(D-3) , 
\e 
\nit where $D=\dim \; \g$.  For $\alga_2$, $X_2$ is the representation
$20=10 + \overline{10}= (3,0)\oplus(0,3)$, a pair of conjugate
irreps. For the exceptional Lie algebras, see Table~\ref{tab_order2}.

\begin{table}
\begin{center}
\begin{tabular}{|c|r|r|r|r|r|} 
\hline
       & $\g_2$ & $\f_4$ & $\alge_6$ & $\alge_7$ & $\alge_8$ \\ 
\hline \hline
$ad$   &  14   &  52  &  78   & 133 & 248  \\ 
       & (10) & (1000) & $(000001)$ & $(1000000)$ & $(00000010)$ \\ \hline 
$X_2$  &  $77^\prime$   &  1274 &  2925  & 8645 & 30380  \\ 
       & (03) & (0100) & $(001000)$ & $(0100000)$ & $(00000100)$ \\ \hline 
$R_1$  &  1    &  1     &  1     & 1        & 1     \\ \hline
$R_2$  &  27    &  324    &  650    & 1539 &  3875  \\ 
       & (02) & (0002) & $(100010)$ & $(0000100)$ & $(10000000)$ \\ \hline
$R_3$  & 77 & 1053  & 2430 & 7371   &  27000  \\
       & (20) & (2000) & $(000002)$ & $(2000000)$ & $(00000020)$    \\ \hline
\end{tabular}
\end{center}
\caption{\label{tab_order2} Irreps of $\g$ for $ad\otimes ad$.}
\end{table}

A universal and important property of the family of irreps $X_2$ is
the result
\be \label{I5} c^{(2)}(X_2)=2. \e \nit
It is of special relevance to the work
described here, because families $X_j$ with
\be \label{I51} c^{(2)}(X_j)=j, \e \nit
for $j \leq 4$ are known to appear in the $j$-th antisymmetric tensor
power of $ad$, and we extend this knowledge beyond $j=4$ here.

The result corresponding to (\ref{I3}) for the symmetric piece of $ad
\otimes ad$ is not universal, but for each $\g \in \cl{F}$ we have a
result of the form
\be \label{I6} (ad \otimes ad)_S = R_1 \oplus R_2 \oplus R_3 , \e \nit 
defining three families of irreps as given in
Table~\ref{tab_order2}. 

These families enjoy a variety of nice properties: for each family we
have single formulas for the dimension of, and for the $\cl{C}^{(2)}$
eigenvalue, of its members as a function of $D=\dim \; \g$. We do not
need these details here. As far as we can determine, the first full
explicit analysis of $ad \otimes ad$ appears in \cite{meyAH}.

The analysis just discussed for $ad \otimes ad$ gives rise to a
natural conjecture --- the Deligne conjecture \cite{del} --- that the
$j$-th tensor powers of $ad$ for the exceptionals possess uniform
decompositions into irreps.  That this does indeed happen, defining
further families of irreps, has been comprehensively confirmed by
algebraic computation in \cite{CdM}, using \cite{LiE}, for $j=3,4$
and established independently of computational procedures in
\cite{lm}.

\begin{table}
\begin{center}
\begin{small}
\begin{tabular}{|c|r|r|r|r|r|r|r|r|}
\hline
           & $\alga_1$ 
           & $\alga_2$ 
           & $\g_2$ 
           & $\algd_4$ 
           & $\f_4$ 
           & $\alge_6$ 
           & $\alge_7$ 
           & $\alge_8$ \\ 
\hline 
\hline
$d_3$      & $-5$ 
           & $0$ 
           & $182$ 
           & $3\cdot840$ 
           & $19448$ 
           & $70070$ 
           & $365750$ 
           & $2450240$ \\ 
\hline
$X_3$      & $(4)$ 
           & ---
           & $(04)$ 
           & $\begin{array}{r}(0022)\cr\oplus(2002)\cr\oplus(2020)\end{array}$ 
           & $(0020)$ 
           & $(010100)$ 
           & $(0010000)$ 
           & $(00001000)$ \\ 
\hline
\hline
$d_4$      & $0$
           & $-35-35$
           & $0$
           & $3\cdot 3675$
           & $205751$ 
           & $2\cdot 600600$  
           & $11316305$  
           & $146325270$  \\ 
\hline
$X_4$      & ---
           & $(14)\oplus(41)$
           & ---
           & $\begin{array}{r}(1013)\cr\oplus(1031)\cr\oplus(3011)\end{array}$
           & (0021) 
           & $\begin{array}{r}(020010)\cr\oplus(100200)\end{array}$
           & $(0001001)$ 
           & $(00010000)$ \\
\hline 
\end{tabular}
\end{small}
\end{center}
\caption{\label{tab_order34} The complete reduction of the
representations $X_3$ and $X_4$ of $\g\in\cl{F}$ in Dynkin coordinates and the
results $d_3(\dim\g)$ and $d_4(\dim\g)$ of the dimension
formulas~(\ref{ab2}) and~(\ref{ab3}).} 
\end{table}

We do not review this in full, but note, for $j=3$, only the result
\be 
\label{ab1} ad^{\wedge 3}=(ad \otimes ad \otimes ad)_A
  = ad \oplus R_2 \oplus R_3 \oplus X_2 \oplus X_3, 
\e
\nit valid for all $\g \in \cl{F}$. The important fact here is that
(\ref{ab1}) defines only one new family beyond those already
understood from the study of $ad \otimes ad$ which we denote by $X_3$
(Table~\ref{tab_order34}). The irreps involved here possess two notable
properties, natural analogues of (\ref{I4}) and (\ref{I5}). Their
dimension is given by a polynomial in $D$,
\begin{equation}
\label{ab2}
 d_3(D) = \fract{1}{3!} D(D-1)(D-8),
\end{equation}
and equation~(\ref{I51}) holds. The factor $D-8$ in~(\ref{ab2})
reflects that fact that $\alga_2$ has no irrep with $c^{(2)}=3$ and
hence no member in the $X_3$ family. For $\alga_1$, the the structure
of~(\ref{ab1}) also collapses, and there is no representation with
$c^{(2)}=3$ in $ad^{\wedge 3}$ either. However, for $D=3$, the
dimension formula (\ref{ab2}) gives the answer $-5$, and $\alga_1$
does indeed have a representation of dimension $5$ with $c^{(2)}=3$
which is, however, not contained in $ad^{\wedge 3}$. No systematic
understanding has been obtained of why negative values of the
dimension formula are often seen to happen and seem to make some sort
of sense. More examples follow in our work for
$j=5,\ldots,9$. Finally, we remark that the three irreps that occur
for $\algd_4$ (Table~\ref{tab_order34}), are related by diagram
automorphisms.

Next we note that analysis of $ad^{\wedge 4}$ brings into the
discussion only one further family of representations beyond those
that were mastered within the discussion (not reviewed here, but see
\cite{CdM}) of $ad^{\otimes3}$.  We denote this family as
$X_4$ (Table~\ref{tab_order34}).

The two irreps, $(100200)$ and $(020010)$, that are listed for
$\alge_6$ are related by diagram automorphisms as are $(0022)$,
$(2002)$ and $(2020)$, for $\algd_4$. The dimension formula
\begin{equation}
\label{ab3}
 d_4(D) = \fract{1}{4!} D(D-1)(D-3)(D-14),
\end{equation}
already indicates that $\alga_1$ and $\g_2$ have no member in the $X_4$
family. Indeed, there do not exist any irreps of $\alga_1$ and $\g_2$
with $c^{(2)}=4$. For $\alga_2$ we have again the phenomenon that
(\ref{ab3}) gives a negative result, here $-70$. Indeed, $\alga_2$ has
got exactly one pair of conjugate irreps with $c^{(2)}=4$ which have dimension
$35+35$. For all irreps listed in Table~\ref{tab_order34}, (\ref{I51})
is satisfied.

The dimension formulas given here in (\ref{I4}),(\ref{ab2}) and
(\ref{ab3}) are equivalent to results given in \cite{del} and
\cite{CdM}, where other parametrisations of family properties are
used: see Sec.~\ref{sect_parameter} below. The $c^{(2)}$ eigenvalues
can also be found in these sources.

%--------------------------------------------------------------------------------------------
\subsection{Summary of new results}
%--------------------------------------------------------------------------------------------
\label{sect_content}

We now turn to to the problem of extending uniformity properties for
$\g \in \cl{F}$ into the case of
$ad^{\otimes5},\ldots,ad^{\otimes9}$. A systematic extension would
seem to entail massive computational effort, but confirmation that the
nice picture known for $j \leq 4$ does not stop at $j=4$ can be
provided.

\begin{table}
\begin{center}
\begin{small}
\begin{tabular}{|c|r|r|r|r|r|r|r|r|} 
\hline
           & $\alga_1$ 
           & $\alga_2$ 
           & $\g_2$ 
           & $\algd_4$ 
           & $\f_4$ 
           & $\alge_6$  
           & $\alge_7$ 
           & $\alge_8$ \\ 
\hline \hline
$d_5$      & $0$
           & $-64$
           & $-924$
           & $\begin{array}{l}3\cdot 3696\cr+15092\end{array}$
           & $\begin{array}{l}629356\cr+952952\end{array}$
           & $\begin{array}{l}2\cdot 1559376\cr+12514788\end{array}$
           & $\begin{array}{l}163601438\cr+109120648\end{array}$
           & 6899079264 \\ 
\hline
$X_5$      & ---
           & $(33)$
           & $(14)$  
           & $\begin{array}{r}(0104)\cr\oplus(0140)\cr\oplus(4100)\cr\oplus(2022)\end{array}$
           & $\begin{array}{r}(0030)\cr\oplus(0103)\end{array}$
           & $\begin{array}{r}(030000)\cr\oplus(000300)\cr\oplus(110110)\end{array}$
           & $\begin{array}{r}(0000102)\cr\oplus(0002000)\end{array}$
           & (00100000) \\
\hline 
\end{tabular}
\end{small}
\end{center}
\caption{\label{tab_order5} Irreps related to (\ref{ab5}).}
\end{table}

Looking at $ad^{\wedge j}$ for $j=2,3,4$ motivates easy but compelling
conjectures. It is natural to expect that there exist, for higher $j$
values, identifiable families $X_j$ of representations of
$\g\in\cl{F}$ occurring in the decomposition of $ad^{\wedge j}$, that
they satisfy~(\ref{I51}) and that
nice dimension formulas exist.

The purpose of this paper then is to attain such knowledge by
confrontation of the cases of $j=5,\ldots,9$. In fact we are able to
provide an identification of the members of families $X_5,\ldots,X_9$
of representations of $\g \in \cl{F}$ that satisfy~(\ref{I51}) and
establish the dimension formulas 
\bea 
 d_5(D) &=& \fract{1}{5!} D(D-3)(D-6)(D^2-21\,D+8),
\label{ab5} \\
 d_6(D) &=& \fract{1}{6!} D(D-1)(D-10)(D^3-34\,D^2+181\,D-144),
\label{ab6} \\ 
 d_7(D) &=& \fract{1}{7!} D(D-2)(D-3)(D-8)(D^3-50\,D^2+529\,D-120),
\label{ab7} \\
 d_8(D) &=& \fract{1}{8!} D(D-1)(D-3)(D-6)(D^4-74\,D^3+1571\,D^2-9994\,D+4200),
\label{ab8} \\
 d_9(D) &=& \fract{1}{9!} D(D-1)(D-3)(D-4)(D-14)(D-26)(D^3-60\,D^2+491\,D-120).
\label{ab9} \ea \nit
We display information in Tables~\ref{tab_order5}
and~\ref{tab_order69} that describe in full the assignments of
representations for the members of the families $X_5,\ldots,X_9$ for
all Lie algebras $\g\in\cl{F}$. There are various features of these
results that need, and will receive, consideration.
\begin{enumerate}
\item 
  The occurrence of the quadratic, cubic and quartic polynomials in
  (\ref{ab5})--(\ref{ab9}) which do not have rational factors.
\item 
  The status of the table entries for $d_j$ for each $\g$ when $j$
  exceeds the first $j$-value $j_0$ for which $d_j$ is not
  positive. For $\alga_2$, $\g_2$, $\algd_4$, $\f_4$ we
  have\footnote{These numbers are related to the highest integer $j$
  for which $ad^{\wedge j}$ contains a Casimir eigenspace of
  eigenvalue $j$~\cite{mal}.} $j_0=3,4,7,10$.
\item 
  The appearance of direct sums of several irreducible representations
  that are not related by diagram automorphisms. This feature is new
  compared with the results of~\cite{CdM}.
\item
  The occurrence of negative values of the dimension formulas
  (\ref{ab5})--(\ref{ab9}).
\item
  The fact that the dimension formulas (\ref{ab5})--(\ref{ab9}) give
  integer results for any integral $D$.
\item
  The question of whether these patterns extend beyond the members of
  the family $\cl{F}$.
\end{enumerate}

The ensuing material is organised as follows. For comparison with the
work of others in Sec.~\ref{sect_parameter}, we mention
parametrisations alternative to $D=\dim \; \g$. In
Sec.~\ref{sect_power5}, we explain our construction of the dimension
formulas (\ref{ab5})-(\ref{ab9}). Sec.~\ref{sect_otheralg} then poses
the obvious question: are the results discussed here for $\g \in
\cl{F}$ universal? The results for $j=2$ are well-known to be
universal in that they apply, not only to $\g \in \cl{F}$, but to all
simple $\g$. To what extent if any does a similar statement hold for
higher $j$? We are unable to give a systematic algebraic analysis of
the situation, but can easily gain some insight into it, by giving an
empirical analysis of the cases of the simple Lie algebras $\algb_2,
\algb_3, \algc_3, \alga_3,\ldots,\alga_5$. Further insight, partially
motivated by the appearance of $D-6$ factors in $d_5(D)$ and $d_8(D)$,
comes from study in Sec.~\ref{sect_a1} of the cases of
$\alga_1\oplus\alga_1$, and the corresponding three-fold and four-fold
direct sum. Sec.~\ref{sect_conclusion} contains a conclusion and a
list of the most obvious open questions.

%--------------------------------------------------------------------------------------------
\subsection{Alternative parametrisations}
%--------------------------------------------------------------------------------------------
\label{sect_parameter}

\begin{table}
\begin{center}
\begin{tabular}{|c|r|r|r|r|r|r|r|r|} 
\hline
           & $\alga_1$ & $\alga_2$ & $\g_2$ & $\algd_4$ & $\f_4$ & $\alge_6$ & $\alge_7$ & $\alge_8$ \\ 
\hline \hline
$D=\dim\g$ & $3$   & $8$   & $14$  & $28$  & $52$  & $78$  & $133$ & $248$ \\ 
\hline
  $\alpha$ & $\frac{1}{2}$ & $\fract{1}{3}$ & $\fract{1}{4}$ &
           $\frac{1}{6}$ & $\fract{1}{9}$ & $\fract{1}{12}$  &
           $\fract{1}{18}$ & $\fract{1}{30}$ \\ 
\hline
    $m$    & $-\frac{4}{3}$ & $-1$ & $-\fract{2}{3}$ & $0$ & $1$ & $2$
           & $4$ & $8$ \\ 
\hline
\end{tabular}
\end{center}
\caption{\label{tab_notation}Parameters $D$, $m$ and
$\alpha$. See~(\ref{P1}) and the text for details.} 
\end{table}

In general in our work, we prefer to give formulas for the dimensions
of and the quadratic Casimir eigenvalues of members of a family of
irreps of the Lie algebras in $\cl{F}$ as a function of the single
parameter $D=\dim \; \g$, but other parameters are used in the
literature. To aid comparison of our discussion with related work in
other sources, we have drawn up Table~\ref{tab_notation}.

The parameter $\alpha$ is used in \cite{del} and \cite{CdM}, while $m$
is used in \cite{lm} and \cite{bw}. The connection between the
different parameters can be obtained from
\bea D & = & \frac{{2(3m+7)(5m+8)}}{m+4} \nonumber \\
\frac{1}{\alpha} & = & 3(m+2) = h^{\vee} , \label{P1} \ea \nit
where $h^{\vee}$ is the dual Coxeter number, (p.~37 of~\cite{FSS}). Also
$\alpha$ is related to the eigenvalue of the Casimir operator of the
members of the $R_3$ family
\be \label{P2} c^{(2)}(R_3)=2(1+\alpha). \e \nit
For the exceptional algebras in the last line of the Freudenthal magic
square, the $m$ values in Table~\ref{tab_notation} have this
interpretation: the division algebra used in their Freudenthal
construction has dimension $m$.

% ============================================================================================
%
\section{The dimension formulas}
%
% ============================================================================================
\label{sect_power5}

\begin{table}
\begin{center}
\begin{tabular}{|c|r|r|r|r|r|} 
\hline
  Root & $\alpha_1$ &  $\alpha_2$ &  $\alpha_3$ &  $\alpha_4$ & Height \\
\hline\hline
$x_{24}$ & $2$ & $3$ & $4$ & $2$ & $11$ \\
$x_{23}$ & $1$ & $3$ & $4$ & $2$ & $10$ \\
$x_{22}$ & $1$ & $2$ & $4$ & $2$ & $9$ \\
$x_{21}$ & $1$ & $2$ & $3$ & $2$ & $8$ \\
$x_{20}$ & $1$ & $2$ & $3$ & $1$ & $7$ \\
$x_{19}$ & $1$ & $2$ & $2$ & $2$ & $7$ \\
$x_{18}$ & $1$ & $2$ & $2$ & $1$ & $6$ \\
$x_{17}$ & $1$ & $1$ & $2$ & $2$ & $6$ \\ 
$x_{16}$ & $0$ & $1$ & $2$ & $2$ & $5$ \\
$x_{15}$ & $1$ & $1$ & $2$ & $1$ & $5$ \\
$x_{14}$ & $1$ & $2$ & $2$ & $0$ & $5$ \\
$\vdots$ &     &     &     &     & $\vdots$ \\
\hline
\end{tabular}
\end{center}
\caption{\label{tab_rootf4}The highest positive roots $x_j$ of $\f_4$
in terms of the simple roots $\alpha_j$ and their height.}
\end{table}

In the study of $ad^{\wedge j}$ up to $j=4$~\cite{CdM} it was
sufficient to identify the irreducible component of the highest weight
in the antisymmetric power $ad^{\wedge j}$ and then to determine the
direct sum $X_j$ of all irreps that can be obtained from the former by
the application of diagram automorphisms. The dimension
formulas~(\ref{ab2}) and~(\ref{ab3}) then agree with the interpolation
polynomial for which $d_j(\dim\g)=\dim X_j$ for all algebras
$\g\in\cl{F}$ for which the corresponding $X_j$
satisfies~(\ref{I51}). In our notation, it is already a non-trivial
fact that a polynomial in $D=\dim\g$ is sufficient to parametrise the
relevant dimensions for all algebras $\g\in\cl{F}$.

At $j=5$, however, the same strategy does not result in any simple
dimension formula at all. The solution is to modify the strategy and
to choose $X_j$ to be the entire Casimir eigenspace of $ad^{\wedge
j}$, \ie\ the maximal $X_j$ that satisfies~(\ref{I51}), or if there is
no non-trivial subspace with this property when the algebra $\g$ has
dropped out of the full picture, to choose a suitable direct sum of
(other) irreps of $\g$ that satisfy~(\ref{I51}). We have arrived at
this result purely empirically, searching for simple and in particular
polynomial dimension formulas, and we have found the following
rule\footnote{We thank J.~Landsberg for bringing the
article~\cite{Kos} to our attention in which the Casimir $j$
eigenspace of $ad^{\wedge j}$ is characterized by an algebraic
condition equivalent to the rule we state here.} which characterizes
the direct summands of $X_j$.

The rule specifies how to select $j$ distinct roots of $\g$ whose sum
is the highest weight of an irrep that is contained in $ad^{\wedge
j}$. Whenever an algebra $\g$ has not yet dropped out of the full
picture (as explained above), the rule finds all irreps that both
occur in $ad^{\wedge j}$ and have $c^{(2)}=j$. If the algebra has
dropped out, the rule finds some other irreps in $ad^{\wedge j}$.

\begin{figure}[t]
\begin{center}
\input{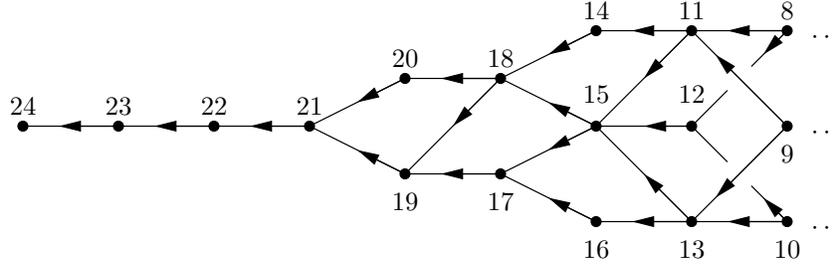}
\end{center}
\caption{\label{fig_roots_f4} 
  A part of the root lattice of $\f_4$ as a directed graph. The
  vertices correspond to the roots $x_k$ and are numbered as in
  Table~\ref{tab_rootf4}.}
\end{figure}

We explain the procedure for $\f_4$ whose roots are given in
Table~\ref{tab_rootf4}. Consider the root lattice of $\f_4$, drawn as
a directed graph in Figure~\ref{fig_roots_f4}. The vertices correspond
to the roots $x_k$ and are numbered as in the table. There is a
directed arrow from $x_k$ to $x_\ell$, denoted by a pair
$(x_k,x_\ell)$, if and only if $x_\ell=x_k+\alpha$ for some simple
root $\alpha$.

Let $X=\{x_1,x_2,\ldots\}$ denote the set of all roots. For
$j=1,2,\ldots$ consider subsets $S\subset X$ of cardinality $|S|=j$. A
subset $S$ is called \emph{admissible} if for each $x\in S$ and each
arrow $(x,y)$ in the graph, we have also $y\in S$.

Observe that in the $\f_4$ example, the highest root $x_{24}$ is
contained in any non-empty admissible subset. Similarly, the second
highest root $x_{23}$ is contained in any admissible set of
cardinality at least $2$, and so on.

\bigskip
\nit{\bf Observation:} 
Given an admissible subset $S$, $|S|=j$, the weight
\begin{equation}
  w := \sum_{x\in S}x
\end{equation}
is the highest weight of an irrep of $\g$ which is contained in
$ad^{\wedge j}$. Any irrep of $c^{(2)}=j$ that occurs in $ad^{\wedge
j}$ can be found from this rule.

\bigskip
Given the root lattices of the algebras $\g\in\cl{F}$, we can use this
rule in order to obtain a list of all irreps that are both contained
in $ad^{\wedge j}$ and also have $c^{(2)}=j$. This information forms
the basis for the higher dimension formulas. 

A point for $\algd_4$ regarding admissible sets is worth noting. For
$\algd_4$, $d_4(28)=3\cdot 3675$ involving a triple of irreps with
$c^{(2)}=4$ which occur in $ad^{\wedge 4}$. But $\algd_4$ also has
$1925=(0300)$ with $c^{(2)}=4$ which is not part of $ad^{\wedge
4}$. If we examine the admissible sets for $\algd_4$ at $j=4$, one
finds sets for the $3675$ dimensional irreps, but not for $1925$.

We obtain the dimension formula~(\ref{ab5}),
\begin{equation}
\label{h2}
 d_5(D) = \fract{1}{5!} D(D-3)(D-6)(D^2-21\,D+8),
\end{equation}
as the interpolation polynomial using the data for six algebras of the
family $\cl{F}$ from Table~\ref{tab_order5}.

If we hoped that the right side of (\ref{h2}) would be the product of
factors linear in $D$, like the formula for $d_j(D)$ for lower $j$
then we have been disappointed. However the expectation was based on
viewing these formulas, as in \cite{lm}, in relation to the Weyl
formula for the dimensions of irreps of Lie algebras, and that view is
valid as long as the families $X_j$ involve only irreps (up to diagram
automorphisms), \ie\ for $j \leq 4$.  But it is not valid for $j=5$
and the assignments for $X_5$ already made, and so the basis for the
hope has gone.

In case it might be thought that the use of the parameter $m$ of
Table~\ref{tab_notation} might improve the status of (\ref{h2}), we
note that (\ref{P1}) implies 
\be 
\label{h101}
D^2-21\,D+8=\frac{6(15m^2+67m+68)(10m^2+27m+28)}{(m+4)^2}, 
\e
in which each one of the quadratic expressions in view has
discriminant $409$ and does not have rational factors.

The dimension formula~(\ref{h2}) gives negative values for $\alga_2$
and $\g_2$ (Table~\ref{tab_order5}), in each case referring to the
unique irrep of the Lie algebra in question with $c^{(2)}=5$.

Although it would not be correct to assign $924$ of $\g_2$ to $X_5$,
since it does not occur in the decomposition for $\g_2$ of $ad^{\wedge
5}$, the situation is similar to the one found for $\alga_2$ at the
previous stage: we found there not a proper member of the $X_4$ family
for $\alga_2$, but one with the correct value of $c^{(2)}$ and the
negative of the correct dimension. Such things are prevalent also in
the work of \cite{del} and \cite{CdM}, but not explained.

As one goes to higher $j$ in an effort to push the search for families
and for dimension formulas for them higher, one expects more and more
algebras to drop out of the full picture much in the way that
$\alga_2$ did beyond $j=2$ and $\g_2$ did beyond $j=3$. This will
happen just because $\dim ad^{\wedge j}$ eventually becomes too small
to contain any irrep of $c^{(2)}=j$. We also expect that for Lie
algebras that have dropped out of the full picture, \ie\ out of
correctly assigning members to families, use of dimension formulas
will continue to yield in modulus representations carrying the correct
Casimir eigenvalue for the family in question.

As we are looking for a dimension formula $d_j(D)$ which is a
polynomial of degree $j$ in $D$, we need $j+1$ Lie algebras to fix its
coefficients and then another Lie algebra in order to confirm that the
dimension formula contains non-trivial information. 

\begin{table}
\begin{center}
\begin{tiny}
\begin{tabular}{|r|r|r|r|r|r|r|r|} 
\hline
           & $\alga_2$ 
           & $\g_2$ 
           & $\algd_4$ 
           & $\f_4$ 
           & $\alge_6$  
           & $\alge_7$ 
           & $\alge_8$ \\ 
\hline \hline
$d_6$      & $\begin{array}{r}28\cr+28\end{array}$
           & $-1547$
           & $3\cdot 1386$
           & $\begin{array}{r}7113106\cr+1850212\end{array}$
           & $\begin{array}{r}2\cdot 47783736\cr+64205141\end{array}$
           & $\begin{array}{r}4260501784\cr+1063409347\end{array}$
           & 267413986840 \\ 
\hline
$X_6$      & $\begin{array}{r}(06)\cr\oplus(60)\end{array}$
           & $(23)$
           & $\begin{array}{r}(0006)\cr\oplus(0060)\cr\oplus(6000)\end{array}$
           & $\begin{array}{r}(0112)\cr\oplus(1005)\end{array}$
           & $\begin{array}{r}(120100)\cr\oplus(010210)\cr\oplus(201020)\end{array}$
           & $\begin{array}{r}(0001101)\cr\oplus(0000013)\end{array}$
           & (01000001) \\
\hline 
\hline
$d_7$      & $0$
           & $\begin{array}{r}1254\cr-748\end{array}$
           & $\begin{array}{r}-3\cdot 23400\cr-114400\end{array}$
           & $\begin{array}{r}1264120\cr+19214624\cr+15997696\end{array}$
           & $\begin{array}{r}2\cdot 466237200\cr+221077350\cr+152423700\end{array}$
           & $\begin{array}{r}2457458575\cr+44406104600\cr+39557939200\end{array}$
           & $\begin{array}{r}5006235840320\cr+3754721200320\end{array}$\\
\hline 
$X_7$      & ---
           & $\begin{array}{r}(07),\cr(40)\end{array}$
           & $\begin{array}{r}(0204)\cr\oplus(0240)\cr\oplus(4200)\cr\oplus(2122)\end{array}$
           & $\begin{array}{r}(0007)\cr\oplus(1014)\cr\oplus(0202)\end{array}$
           & $\begin{array}{r}(101120)\cr\oplus(211010)\cr\oplus(020200)\cr\oplus(300031)\end{array}$
           & $\begin{array}{r}(0000004)\cr\oplus(0001012)\cr\oplus(0010200)\end{array}$
           & $\begin{array}{r}(10000002)\cr\oplus(02000000)\end{array}$\\
\hline
\hline
$d_8$      & $-125$
           & $3003$
           & $\begin{array}{r}-3\cdot 15015\cr-3\cdot 169785\end{array}$
           & $\begin{array}{r}65609375\cr+15611882\cr+13530946\end{array}$
           & $\begin{array}{r}3863940795\cr+2\cdot 764156250\cr+2\cdot
                              1533061530\cr+133024320\end{array}$
           & $\begin{array}{r}135058673750\cr+848520798125\cr+204501797500\end{array}$
           & $\begin{array}{r}35361935272950\cr+212182409960235\end{array}$\\
\hline 
$X_8$      & $(44)$
           & $(16)$
           & $\begin{array}{r}(0106)\cr\oplus(0160)\cr\oplus(6100)\cr\oplus(1213)\cr\oplus(1231)\cr\oplus(3211)\end{array}$
           & $\begin{array}{r}(1104)\cr\oplus(0016)\cr\oplus(0300)\end{array}$
           & $\begin{array}{r}(111110)\cr\oplus(002030)\cr\oplus(302000)\cr\oplus(200131)\cr\oplus(310021)\cr\oplus(400040)\end{array}$
           & $\begin{array}{r}(0001003)\cr\oplus(0010111)\cr\oplus(0100300)\end{array}$
           & $\begin{array}{r}(00000003)\cr\oplus(11000001)\end{array}$\\
\hline
\hline
$d_9$      & $\begin{array}{r}80\cr+80\end{array}$
           & $0$
           & $\begin{array}{r}-3\cdot 215600\cr+245700\end{array}$
           & $64194312$
           & $\begin{array}{r}2\cdot 1621233900\cr+2\cdot 5284021600\cr+2\cdot7587880300\cr+16162889600\end{array}$
           & $\begin{array}{r}3581756027850\cr+3334268437500\cr+6557368727910\cr+539884745400\end{array}$
           & $\begin{array}{r}2624940724551600\cr+3500209714601600\end{array}$\\
\hline 
$X_9$      & $\begin{array}{r}(17)\cr\oplus(71)\end{array}$
           & ---
           & $\begin{array}{r}(0322)\cr\oplus(2302)\cr\oplus(2320),\cr(3033)\end{array}$
           & $(0106)$
           & $\begin{array}{r}(300140)\cr\oplus(410030)\cr\oplus(101041)\cr\oplus(401011)\cr
                              \oplus(012020)\cr\oplus(202100)\cr\oplus(210121)\end{array}$
           & $\begin{array}{r}(0010102)\cr\oplus(0020020)\cr\oplus(0100211)\cr\oplus(1000400)\end{array}$
           & $\begin{array}{r}(01000002)\cr\oplus(20100000)\end{array}$\\
\hline
\end{tabular}
\end{tiny}
\end{center}
\caption{\label{tab_order69} Irreps related to (\ref{ab6})--(\ref{ab9}).}
\end{table}

Table~\ref{tab_order69} shows the values $d_j(D)$ of the dimension
formulas~(\ref{ab6})--(\ref{ab9}), $j=6,\ldots,9$ and the assignment
of representations $X_j$. For $\alga_1$, $d_6=-7$ and the seven
dimensional irrep of $\alga_1$ has $c^{(2)}=6$, but $d_7=d_8=d_9=0$ as
expected.

For $j=6,7$, we have obtained the dimension formulas~(\ref{ab5})
and~(\ref{ab6}) as the interpolation polynomial for the eight values
$d_j(D)$ where $D=\dim\g$ for the eight Lie algebras
$\g\in\cl{F}$. Whenever $ad^{\wedge j}$ contains irreps of
$c^{(2)}=j$, then we choose $X_j$ to be their direct sum and
$d_j(\dim\g)=\dim X_j$. Whenever $ad^{\wedge j}$ does not contain any
irreps of $c^{(2)}=j$, so that $\g$ has dropped out of the full
picture, then $d_j(\dim\g)$ is the sum or difference of the dimensions
of all irreps of $\g$ with $c^{(2)}=j$. In this case, the signs of the
summands are chosen by trial and error so that we obtain a `simple'
dimension formula, \ie\ one which has as many linear factors as
possible, which has only `small' coefficients and for which $d_j(D)$
is an integer for any integral $D$. By experimentation with these
interpolations, we always find a unique choice of signs which
dramatically, absolutely dramatically, simplifies the interpolation
polynomial.

For $j=7$, it is of course a trivial fact that we can use the data of
$8$ algebras in order to uniquely fix a polynomial $d_j(D)$ of degree
$7$. The simplicity of the resulting formula~(\ref{ab7}) is, however,
a highly non-trivial property.

Comparing the dimension formulas~(\ref{ab3})--(\ref{ab7}), the
following general pattern emerges: We have $d_j(0)=0$, and the leading
term is $\frac{1}{j!}D^j$. For $j=8$, we now assume these two
conditions and can therefore determine a polynomial of degree eight
from only seven additional data points. We employ all
algebras $\g\in\cl{F}$ except for $\algd_4$ and obtain~(\ref{ab8}).

For $\algd_4$, we discover the following exception from the rules
stated so far. The dimension formula~(\ref{ab8}) yields
$d_8(28)=-554400$. There exist indeed representations of $\algd_4$
with $c^{(2)}=8$, namely $(0106)\oplus(0160)\oplus(6100)$ of dimension
$3\cdot 15015$ and $(1213)\oplus(1231)\oplus(3211)$ of dimension
$3\cdot 169785$, and indeed $d_8(28)$ is the negative sum of their
dimensions. However, $\algd_4$ has got further representations with
$c^{(2)}=8$ that do not play any role in the dimension formula
$d_8(D)$, namely $(0044)\oplus(4004)\oplus(4040)$ with dimension
$3\cdot 35035$.

For $j=9$, we again assume the two conditions and determine a
polynomial of degree 9 from eight data points, making use of all eight
algebras $\g\in\cl{F}$.

% ============================================================================================
%
\section{Dimension formulas for simple $\g$ not in $\cl{F}$}
%
% ============================================================================================
\label{sect_otheralg}

\begin{table}
\begin{small}
\begin{center}
\begin{tabular}{|c|c|c|c|c|c|c|}
\hline
         & $\algb_2$ & $\algb_3$ & $\algc_3$ & $\alga_3$ & $\alga_4$ & $\alga_5$ \\ 
\hline 
\hline
   $d_2$  &  $35$   &  $189$  &  $189$  & $45+{\overline {45}}$ & 
             $126+{\overline {126}}$ &  $280+{\overline {280}} $ \\ 
\hline
   $X_2$  &  $(12)$ & $(101)$ & $(210)$ & 
             $\begin{array}{c}(210)\cr\oplus(012)\end{array}$ &
             $\begin{array}{c}(5000)\oplus(0005)\end{array}$ &
             $\begin{array}{c}(20010)\oplus(01002)\end{array}$ \\ 
\hline 
\hline
  $d_3$   &  $30$  &  
             $\begin{array}{c}294\cr+616\end{array}$ &
             $\begin{array}{c}385\cr+525\end{array}$ &  
             $\begin{array}{c}35+{\overline {35}}\cr+175\end{array}$&
             $\begin{array}{c}1024\cr+224+{\overline{224}}\end{array}$ &  
             $\begin{array}{c}3675\cr+840+{\overline{840}}\end{array}$ \\ 
\hline  
  $X_3$   &  $(30)$ &  
             $\begin{array}{c}(004)\cr\oplus(202)\end{array}$&
             $\begin{array}{c}(030)\cr\oplus(301)\end{array}$&
             $\begin{array}{c}(400)\cr\oplus(004)\cr\oplus(121)\end{array}$&
             $\begin{array}{c}(1111)\cr\oplus(3010)\oplus(0103)\end{array}$ &
             $\begin{array}{c}(11011)\cr\oplus(30100)\oplus(00103)\end{array}$\\
\hline 
\hline 
  $d_4$   &  $-105$ & 
             $\begin{array}{c}1386\cr+819\end{array}$ &  
             $2205$ & $105$ &    
             $\begin{array}{c}1701+{\overline {1701}}\cr+1176\cr+126+{\overline{126}}\end{array}$ &
             $\begin{array}{c}12250+{\overline{12250}}\cr+6720\cr+1050+{\overline{1050}}\end{array}$ \\ 
\hline
  $X_4$   &  $(14)$ &  
             $\begin{array}{c}(104)\cr\oplus(310)\end{array}$ &
             $(121)$ & 
             $(040)$ & 
             $\begin{array}{c}(2201)\oplus(1022)\cr\oplus(0220)\cr\oplus(5000)\oplus(0005)\end{array}$ & 
             $\begin{array}{c}(21101)\oplus(10112)\cr\oplus(02020)\cr\oplus(41000)\oplus(00014)\end{array}$ \\ 
\hline
\hline 
  $d_5$   &  $\begin{array}{c}-84\cr-154\end{array}$ &
             $378$ &  
             $\begin{array}{c}2457\cr-2079\end{array}$ &
             $\begin{array}{c}-189\cr-{\overline{189}}\cr-729\end{array}$ &
             $\begin{array}{c}3024\cr+{\overline{3024}}\end{array}$ &
             $\begin{array}{c}36750+{\overline{36750}}\cr+34496\cr+12936+{\overline{12936}}\cr
                              +462+{\overline{462}}\end{array}$ \\ 
\hline
  $X_5$    & $\begin{array}{c}(06)\cr\oplus(32)\end{array}$ & 
             $(500)$ & 
             $\begin{array}{c}(022),\cr(501)\end{array}$ &
             $\begin{array}{c}(501)\cr\oplus(105)\cr\oplus(222)\end{array}$ &  
             $\begin{array}{c}(1310)\cr\oplus(0131)\end{array}$ & 
             $\begin{array}{c}(01121)\oplus(12110)\cr\oplus(20202)\cr\oplus(32001)\oplus(10023)\cr\oplus(60000)
                              \oplus(00006)\end{array}$ \\ 
\hline 
\hline
  $d_6$    & --- &  
             $\begin{array}{c}-9009\cr-4312\end{array}$ &
             $\begin{array}{c}-11319\cr-3003\cr+1001\end{array}$  &  
             $\begin{array}{c}-735\cr-875\cr-{\overline{875}}\end{array}$ &
             $\begin{array}{c}-8624-\overline{8624}\cr-924-\overline{924}\cr+1176+\overline{1176}\end{array}$&
             $\begin{array}{c}169785\cr+43120+{\overline{43120}}\cr+25200+{\overline{25200}}\end{array}$ \\ 
\hline 
  $X_6$    & ---  & 
             $\begin{array}{c}(114)\cr\oplus(320)\end{array}$ &
             $\begin{array}{c}(321),\cr(610),\cr(004)\end{array}$ &
             $\begin{array}{c}(141)\cr\oplus(412)\cr\oplus(214)\end{array}$ & 
             $\begin{array}{c}(3202)\oplus(2023)\cr\oplus(6001)\oplus(1006)\cr\oplus(0500)\oplus(0050)\end{array}$&
             $\begin{array}{c}(11211)\cr\oplus(23010)\oplus(01032)\cr\oplus(03200)\oplus(00230)\end{array}$ \\ 
\hline
\end{tabular}
\end{center}
\end{small}
\caption{\label{tab_otheralg}The dimension formulas $d_j(D)$ for some
algebras $\g$ and the corresponding representations $X_j$ in highest
weight notation.} 
\end{table}

The formulas~(\ref{I4}), (\ref{ab2}), (\ref{ab3}),
(\ref{ab5})--(\ref{ab9}) have been derived and discussed in the
context of the extended family $\cl{F}$ of simple Lie algebras $\g$
that include the exceptionals.

It is known however that (\ref{I4}) is universal: a well-defined
representation $X_2$ of each simple $\g$ has its dimension given by
(\ref{I4}) and the eigenvalue $c^{(2)}(X_2)=2$ of its quadratic
Casimir operator. It is natural to ask if the other dimension formulas
(\ref{ab2}), (\ref{ab3}), (\ref{ab5})--(\ref{ab9}) are likewise
universal, and, if not, what, if anything, they can tell us for $\g
\notin \cl{F}$.

No systematic algebraic approach is available, but a large body of
data can readily be assembled, \eg\ using our programs, MAPLE, and
references such as~\cite{cor,slan}. Some indication of the
limitations, if any, of the applicability of (\ref{ab2})--(\ref{ab9})
to $\g \notin \cl{F}$ can certainly be gained by reference to the
cases of $\algb_2 (\cong\algc_2)$, $\algb_3$, $\algc_3$,
$\alga_3,\ldots,\alga_5$.

The entries of Table~\ref{tab_otheralg} for each $\g$ and each $X_j$
show representations of $\g$ --- often direct sums of irreps --- all
with the $c^{(2)}(X_j)=j$. The dimension formulas $d_j(D)$,
$D=\dim\g$, yield sums or differences of the dimensions of the
irreducible components of $X_j$. Note that the feature that $d_j(D)$
gives a difference of the dimensions of irreps was for $\g_2$ first
encountered at $j=7$.

If we extend Table~\ref{tab_otheralg} to $\algb_4$ and $\algc_4$, we
encounter at $j=5$ the same exception from the rules that we have
already seen for $\algd_4$ at $j=8$, namely that the algebra has
already dropped out of the full picture (as explained above), and
there exist many irreps with $c^{(2)}=j$ only some of which are
relevant for the dimension formula.

It is a striking observation that the simple Lie algebras of
Table~\ref{tab_otheralg} all fit into the general pattern. In
particular, the fact that $\algb_2$ does not have any irrep of
$c^{(2)}=6$ can be seen as an `explanation' of the linear factor
$D-10$ in~(\ref{ab6}). It is then a natural question to ask which are
the Lie algebras of dimension $1$, $2$, $4$, $6$ and $26$ that cause
the other integer roots of the dimension
formulas~(\ref{ab5})--(\ref{ab9}).

% ============================================================================================
%
\section{Some further studies}
%
% ============================================================================================
\label{sect_a1}

%--------------------------------------------------------------------------------------------
\subsection{The factor $(D-6)$ in $d_5(D)$ and $d_8(D)$}
%--------------------------------------------------------------------------------------------

To account for the presence of the factors $(D-6)$, consider the case
of $\g=\alga_1\oplus\alga_1$, employing the Cartan matrix 
\be \label{fs1} A= \left( \begin{array}{cc} 2 & 0 \\ 0 & 2 \end{array} \right),
 \e \nit
and a Cartan-Killing form with no relative scaling of the two
$\alga_1$ summands, so that the algebra has an $S_2$ group of diagram
automorphisms.

Let $(j,k)$ denote the irrep of dimension $(j+1)(k+1)$, so that
$ad=(2,0) \oplus (0,2)$. We list in Table~\ref{tab_2a2} irreps of
$\alga_1\oplus\alga_1$ with integer eigenvalues $n$ of the quadratic
Casimir operator, with $d_n(6)$ alongside for comparisons of the type
made systematically made in previous cases.

\begin{table}
\begin{center}
\begin{tabular}{|c|c|c|}
\hline
$c^{(2)}=n$ & irreps & $d_n(6)$ \\ 
\hline \hline
 1 &  $(2,0) \oplus (0,2)$ & $6$ \\
 2 &  $(2,2)$ & $9$ \\
 3 &  $(4,0) \oplus (0,4)$ & $-10$ \\
 4 &  $(4,2) \oplus (2,4)$ & $-30$ \\
 5 & none & 0 \\
 6 &  $(6,0) \oplus (0,6), \; (4,4) $ & $11$ \\
 7 &  $(6,2) \oplus (2,6)$ & $-42$ \\
 8 & none & 0 \\
 9 &  $(6,4) \oplus (4,6)$ & $70$ \\
\hline
\end{tabular} 
\end{center}
\caption{\label{tab_2a2} Data for $\alga_1\oplus\alga_1$.}
\end{table}

The entries for $c^{(2)}=5$ and $c^{(2)}=8$ explain the $(D-6)$ factors in $d_5(D)$ and
$d_8(D)$, and all the other entries follow precisely a now familiar pattern.
Only one entry needs any comment:
\be \label{fs2} d_6(6)=11=25-2\cdot 7, \e \nit
where $25=\dim(4,4)$, $7=\dim(6,0)=\dim(0,6)$

We note also that all the irreps that feature here are either self-conjugate 
or else occur as conjugate pairs, as the $S_2$ invariance of $ad$ requires.

%--------------------------------------------------------------------------------------------
\subsection{$\alga_1\oplus\alga_1\oplus\alga_1$}
%--------------------------------------------------------------------------------------------

In this case we use as Cartan matrix twice the unit matrix, again with
no relative scales, so that the algebra has a group $S_3$ of diagram
automorphisms. Table~\ref{tab_3a2} displays data about all the irreps
with integral values of the Casimir operator. The notation $(j,k,l)$
denotes the irrep with dimension $(j+1)(k+1)(l+1)$, so that
$ad=(2,0,0) \oplus (0,2,0) \oplus (0,0,2)$. To keep the displays as
brief as is reasonable, the notation $r\cdot(a,b,c)$ denotes the
direct sum of all $r$ distinct permutations of $(a,b,c)$. In view of
the automorphism group $S_3$, we may have $r=3$ and $r=6$. Again we
can check that all the data conforms to the expected pattern.

\begin{table}
\begin{center}
\begin{tabular}{|c|c|c|} 
\hline
$c^{(2)}=n$ & irreps & $d_n(9)$ \\ 
\hline \hline
 3 &  $(2,2,2),\;  3\cdot(4,0,0) $ & $12=27-3\cdot5$ \\
 4 &  $6\cdot(4,2,0)$ & $-90=-6\cdot15$ \\
 5 & $ 3\cdot(4,2,2)$ & $-135=-3\cdot45$\\
 6 &  $3\cdot(4,4,0), \; 3\cdot(6,0,0)$ & $54=3\cdot25-3\cdot7$ \\
 7 &  $6\cdot(6,2,0), \; 3\cdot(4,4,2)$ & $3\cdot75-6\cdot21$ \\
 8 & $3\cdot(6,2,2)$ & $-189=-3\cdot63$ \\
 9 &  $6\cdot(6,4,0), \; (4,4,4)$ & $85=6\cdot35-125$ \\ \hline 
\end{tabular} 
\end{center}
\caption{\label{tab_3a2} Data for $\alga_1\oplus\alga_1\oplus\alga_1$.}
\end{table}

%--------------------------------------------------------------------------------------------
\subsection{$\alga_1\oplus\alga_1\oplus\alga_1\oplus\alga_1$}
%--------------------------------------------------------------------------------------------

This example was treated to see one further automorphism group at work. But few
surprises were expected or found. Everything is in full accord with 
expectation. We do not display the data that would make a table like 
Tables~\ref{tab_2a2} and~\ref{tab_3a2}, but note only the situation for $d_9(12)$.
The set of irreps, in notation similar to that used in previous subsections,
that have $c^{(2)}=9$ is
\be \label{fs31} 12\cdot(6,4,0,0), \quad 4\cdot( 4,4,4,0), \quad 4\cdot(7,1,1,1),
\quad 4\cdot(6,2,2,2), \e \nit
with dimensions $12\cdot35, \; 4\cdot125, \; 4\cdot64, \; 4\cdot189$,
and we have
\be \label{fs32} d_9(12)=-836=12\cdot35-4\cdot(189+125). \e \nit
We note that the resolution (\ref{fs32}) does not employ the irreps
$4\cdot(7,1,1,1)$, but such an omission is also familiar in previous cases.
There is some curious numerology in the $n=9$ case: $64=(189-125)$, and similar
things are seen for lower $n$ cases. 

But we leave the analysis here, without including much further data with
features that are of a qualitatively similar nature to what has been
presented. The fact however is that everything follows a coherent if 
far from understood pattern. 

% ============================================================================================
%
\section{Conclusion}
%
% ============================================================================================
\label{sect_conclusion}

We have extended the dimension formulas of~\cite{CdM} for a particular
family up to the ninth power of the adjoint representation. The
formulas~(\ref{ab5})--(\ref{ab9}) describe a further striking
uniformity of the Lie algebras of the exceptional series and our
results of Sec.~\ref{sect_otheralg} and~\ref{sect_a1} indicate even a
uniformity beyond that. The formulas were obtained by inspection of a
large amount of data from tables and from computer calculations and
finally by a considerable amount of trial and error until we had found
the appropriate rules that give rise to `simple' formulas. The fact
that `easy' formulas such as~(\ref{ab5})--(\ref{ab9}) with
coefficients smaller than a few thousand give rise to integers of up
to sixteen digits which precisely correspond to dimensions of
representations of the exceptional Lie algebras, deserves to be seen
as of real significance.

We conclude by listing some particular observations and first ideas
that come to mind.
\begin{itemize}
\item
  The formulas~(\ref{ab5})--(\ref{ab9}) for $d_j(D)$ are polynomials
  in $D=\dim\g$. From~\cite{CdM}, we might have expected only rational
  functions in $\alpha$, $m$ or $D$. In particular, $\dim Y_j$ in the
  notation of~\cite{CdM} ($Y_j$ is the highest weight component of the
  $j$-th totally symmetric tensor power of $ad$) is not a polynomial
  in $D$.
\item
  With the rational functions of~\cite{CdM}, one can search for those
  values of the parameter $\alpha$ for which the result is an integer
  and thus obtain a list of all algebras that conform to the family
  pattern. In our case, however, the formulas for $d_j(D)$ give
  integer results for any integral $D$.
\item
  Whenever the dimension formula ceases to give a strictly positive
  answer and $ad^{\wedge j}$ does not contain any representation of
  the desired $c^{(2)}$, we can successfully describe the
  phenomenology of the situation, but do not have a satisfactory
  explanation of why it occurs.
\item
  The leading coefficient of the formula for $d_j(D)$ is $1/j!$. This
  looks like a growth rate of the dimension $d_j(D)$ of the family
  member $X_j$ if the dimension $D$ of the underlying Lie algebra $\g$
  tends to infinity.
\item
  Our data confirm that the dimension formulas extend to other simple
  Lie algebras not in the family $\cl{F}$ and beyond that also to some
  non-simple Lie algebras. A particularly strong indication for this
  is the fact that the algebras $\algb_2$ and $\alga_1\oplus\alga_1$
  `explain' some of the integer roots of the polynomials~(\ref{ab5}),
  (\ref{ab6}) and~(\ref{ab8}).
\end{itemize}

It is an interesting question of whether one can identify for each
linear factor $(D-m)$ of the dimension formulas $d_j(D)$ a Lie algebra
of dimension $m$ for which there exists no irrep with $c^{(2)}=j$.
For large $j$, however, there is hardly any simple Lie algebra other
than $\alga_1$. It is is therefore crucial to go beyond simple Lie
algebras and to include more examples in order to prove or disprove
the conjecture. In this context, it is a striking observation that
$d_9(D)$ of~(\ref{ab9}) has so many linear factors.

%------------------------------------------------------------------------------
\section*{Acknowledgments}
%------------------------------------------------------------------------------

The research of A.J.M.\ is partly supported by PPARC. H.P.\ is
grateful to Emmanuel College, Cambridge, for a Research Fellowship. We
would like to thank Bruce Westbury for stimulating discussions, for a
preliminary version of the manuscript~\cite{bw} and for copies of
other manuscripts of his research work.

\end{document}